\documentclass[twocolumn,showpacs,superscriptaddress,nofootinbib]{revtex4-2}

\usepackage[latin9]{inputenc}
\usepackage{amsmath}
\usepackage{amssymb}
\usepackage{amsthm}
\usepackage{graphicx}
\usepackage{color}
\usepackage{hyperref}
\hypersetup{%
    pdfborder = {0 0 0}
}

\theoremstyle{plain}
\newtheorem*{theorem*}{Theorem}

\makeatletter

\@ifundefined{textcolor}{}
{
 \definecolor{BLACK}{gray}{0}
 \definecolor{WHITE}{gray}{1}
 \definecolor{RED}{rgb}{1,0,0}
 \definecolor{GREEN}{rgb}{0,1,0}
 \definecolor{BLUE}{rgb}{0,0,1}
 \definecolor{CYAN}{cmyk}{1,0,0,0}
 \definecolor{MAGENTA}{cmyk}{0,1,0,0}
 \definecolor{YELLOW}{cmyk}{0,0,1,0}
}

\usepackage{color}
\usepackage{amsfonts}
\usepackage{graphics}
\usepackage{bm}
\usepackage{epsfig}\usepackage{amsthm}

\newcommand\blfootnote[1]{%
  \begingroup
  \renewcommand\thefootnote{}\footnote{#1}%
  \addtocounter{footnote}{-1}%
  \endgroup
}

\def\identity{\leavevmode\hbox{\small1\kern-3.8pt\normalsize1}}

\renewcommand{\epsilon}{\varepsilon}

\makeatother

\begin{document}
\title{Beating the classical phase precision limit using a quantum neuromorphic platform}

\author{Tanjung Krisnanda}
\affiliation{School of Physical and Mathematical Sciences, Nanyang Technological University, 637371 Singapore, Singapore}
\blfootnote{Corresponding authors: T.K. (tanjung.krisnanda@ntu.edu.sg) or T.C.H.L. (TimothyLiew@ntu.edu.sg)}

\author{Sanjib Ghosh}
\affiliation{School of Physical and Mathematical Sciences, Nanyang Technological University, 637371 Singapore, Singapore}

\author{Tomasz Paterek}
\affiliation{Institute of Theoretical Physics and Astrophysics, Faculty of Mathematics, Physics and Informatics, University of Gda\'{n}sk, 80-308 Gda\'{n}sk, Poland}

\author{Wies\l aw Laskowski}
\affiliation{Institute of Theoretical Physics and Astrophysics, Faculty of Mathematics, Physics and Informatics, University of Gda\'{n}sk, 80-308 Gda\'{n}sk, Poland}
\affiliation{International Centre for Theory of Quantum Technologies, University of Gda\'{n}sk, 80-308 Gda\'{n}sk, Poland}

\author{Timothy C. H. Liew}
\affiliation{School of Physical and Mathematical Sciences, Nanyang Technological University, 637371 Singapore, Singapore}
\affiliation{MajuLab, International Joint Research Unit UMI 3654, CNRS, Universit\'{e} C\^{o}te d'Azur, Sorbonne Universit\'{e}, National University of Singapore, Nanyang Technological University, Singapore}

\begin{abstract}
Phase measurement constitutes a key task in many fields of science, both in the classical and quantum regime. 
The higher precision of such measurement offers significant advances, and can also be utilised to achieve finer estimates for quantities such as distance, the gravitational constant, electromagnetic field amplitude, etc.
Here we theoretically model the use of a quantum network, composed of a randomly coupled set of two-level systems, as a processing device for phase measurement.
An incoming resource state carrying the phase information interacts with the quantum network, whose emission is trained to produce a desired output signal.
We demonstrate phase precision scaling following the standard quantum limit, the Heisenberg limit, and beyond.
This can be achieved using quantum resource states such as NOON states or other entangled states, however, we also find that classically correlated mixtures of states are alone sufficient, provided that they exhibit quantum coherence.
Our proposed setup does not require conditional measurements, and is compatible with many different types of coupling between the quantum network and the phase encoding state, hence making it attractive to a wide range of possible physical implementations.
\begin{flushleft}
\noindent {\bf Keywords}: Phase estimation, standard quantum limit, Heisenberg limit, quantum neural network, quantum reservoir processing, quantum metrology.\\
\end{flushleft}
\end{abstract}

\maketitle

\section{introduction}
Superior precision in measurements is of great value in ample situations, across many scientific disciplines. 
For the case of optical phase measurements, conventional methods involve the use of classical light (laser), passing through a material (sample) after which the signal undergoes interference with a reference, resulting in an output containing phase information from the sample.
However, this method can be detrimental, especially for sensitive samples with a large number of penetrating photons, which can alter their phase.
Furthermore, for samples with fluctuating phases, a method capable of performing rapid phase estimation is desired.
In this case, the laser power limits the number of photons emitted per unit time. 
Consequently the race is towards performing phase estimation with limited number of photons and obtaining high precision output, the regime of which quantum systems offer supremacy over classical counterparts.

In this direction, the classical limitation for the precision of phase $\phi$ is given by the so-called standard quantum limit (SQL), which lower bounds the error $\Delta \phi$ as in the central limit theorem, i.e., $\Delta \phi \propto 1/\sqrt{N}$, where $N$ is the number of particles used in the measurement. 
To overcome this, quantum states possessing quantum properties such as entanglement have been proposed as a resource to obtain precision below the SQL and reach a scaling $\Delta \phi \propto 1/N$, i.e., the so-called Heisenberg limit (HL)~\cite{giovannetti2004quantum}.
Going beyond the SQL has been demonstrated in a number of experiments, for example: optical phase measurements~\cite{nagata2007beating,higgins2007entanglement,slussarenko2017unconditional,daryanoosh2018experimental}; matter phase~\cite{gross2010nonlinear,lucke2011twin}; sensing of a single ion mechanical oscillator~\cite{mccormick2019quantum}; and for magnetic field~\cite{jones2009magnetic,napolitano2011interaction,kong2020measurement} as well as electric field sensing~\cite{facon2016sensitive}.
It is also expected to enhance the detection of gravitational waves at the LIGO~\cite{yu2020quantum}.

A phase estimation setup overcoming the SQL, conventionally involves a Mach-Zehnder interferometer or an improved version~\cite{giovannetti2004quantum,nagata2007beating,higgins2007entanglement,gross2010nonlinear,lucke2011twin,slussarenko2017unconditional,daryanoosh2018experimental}, from which the output retrieves the phase information. 
One of the forms of resource state known as a NOON state was proven useful, i.e., $(|N0\rangle +|0N\rangle)/\sqrt{2}$, where $N$ denotes the number of excitations.
Earlier work utilised these states to demonstrate phase super resolution, corresponding to interference oscillation $N$ times that of single photon resource states~\cite{walther2004broglie,mitchell2004super,resch2007time}.
Super sensitivity, which is associated with phase precision beating the SQL, was reported for $N=2$~\cite{slussarenko2017unconditional}, $N=4$~\cite{nagata2007beating}, and approaching the HL for $N=3$~\cite{daryanoosh2018experimental}.
However, following this method, going towards the HL for higher $N$-NOON states comes with an increasing complexity, requiring conditional measurements.

Recently, neural networks have been fruitful for solving complex problems in a number of fields~\cite{webb2018deep,jones2019setting,topol2019high,hannun2019cardiologist,nagy2019variational,hartmann2019neural,vicentini2019variational,mehta2019high,montavon2012neural,wetzstein2020inference}.
In general, one has an input, a trained network composed of connected nodes that acts as a processing device, and an output. 
Among the different forms of neural network architectures, reservoir computing transpires as one of the competitive candidates~\cite{vandoorne2014experimental,nguimdo2015simultaneous,van2017advances,brunner2019photonic,tanaka2019recent,ballarini2020polaritonic,rafayelyan2020large,dash2020explicit,shi2020approach,przyczyna2020reservoir}, particularly for direct hardware implementation.
Reservoir computing does not require training over the internal connections between network nodes, rather, it is performed only on a single output layer, in which the signals from the network are processed to produce a desired output.
The vast progress of this field has also reached the quantum regime -- termed quantum reservoir computing or processing.
In this case, a quantum network is utilised to execute classical tasks outperforming classical networks~\cite{fujii2017harnessing,govia2021quantum,kalfus2021neuromorphic,xu2021superpolynomial} and also quantum information tasks, such as: state characterisation~\cite{ghosh2019quantum,ghosh2020reconstructing}; preparation of quantum resource states~\cite{ghosh2019quantum2,creating}; and a platform for quantum computing~\cite{ghosh2020universal}.
See also Refs.~\cite{markovic2020quantum,ghoshquantum} for comparison of different implementations.

Motivated by the direction towards high precision measurements and the vastly growing field of neural networks, here we demonstrate theoretically the use of a quantum network as a processing device for high precision phase measurement. 
In particular, a resource state carrying phase information after passing through a sample acts as an input, which then interacts with a quantum network (QN) composed of randomly coupled two-level quantum systems, which we refer to as network nodes. 
The emission from the network (or the measured occupations of the network nodes) is then linearly combined through an output layer, which is trained using ridge regression, in order to generate a desired output signal. 
We show that our method can perform phase estimation with precision adhering to the SQL, HL, and beyond.
We also show that higher QN size offers improved precision and that one can utilise time-integrated measurements for the emission from the QN before combination through an output layer, which is experimentally friendly.
We also discuss different types of noise that may affect the QN nodes.
Our method is applicable for different types of coupling involved between the input and the network, thus making it desirable for a wide range of physical implementations.
Last, we compare our results to the quantum Cram\'{e}r-Rao bound and show that for some QN parameters, one may obtain near-saturation phase estimation errors.

\section{The setup}

Here we consider a generic simple model and note that our treatment can potentially be applied to physical systems for experimental realisations, such as: randomly coupled quantum dots~\cite{alivisatos1996semiconductor,bimberg1999quantum}; arrays of atomic systems~\cite{esslinger2010fermi,hofstetter2018quantum,tarruell2018quantum,chang2014quantum,poshakinskiy2020quantum,browaeys2020many}; photonic modes in connected optical resonators~\cite{carusotto2009fermionized,bardyn2012majorana,vaneph2018observation} or coupled waveguides~\cite{moughames2020three}; 
interacting exciton-polariton systems~\cite{angelakis2017quantum,delteil2019towards,emmanuele2020highly,kyriienko2020nonlinear}; superconducting qubits~\cite{haroche2020cavity,blais2020quantum,clerk2020hybrid,carusotto2020photonic}; and programmable QN with a multimode fibre~\cite{leedumrongwatthanakun2020programmable}.

\begin{figure}[h]
\centering
\includegraphics[width=0.48\textwidth]{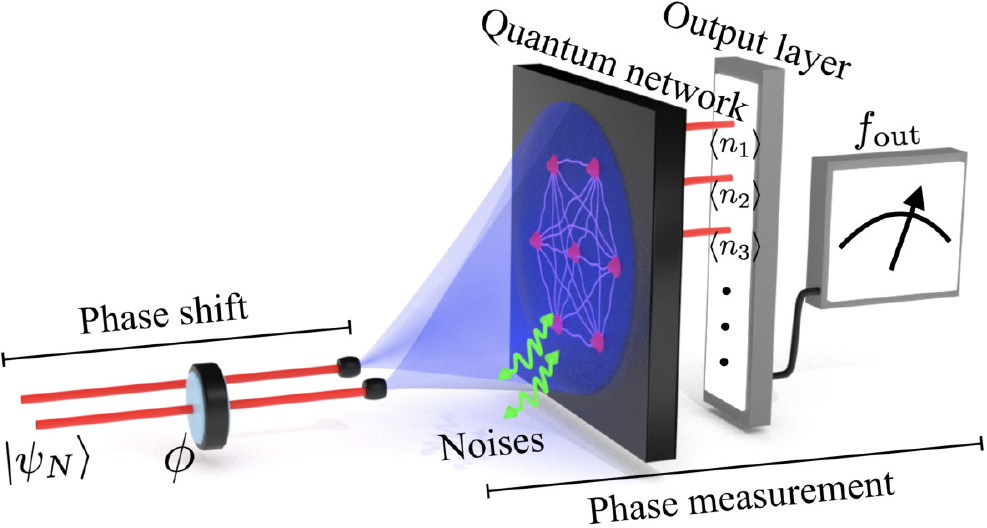}
\caption{Setup for high precision phase measurement. A quantum resource state $|\psi_N \rangle$ picks up a phase, after which it interacts with a quantum network consisting of randomly coupled two-level nodes. The phase information is embedded in the QN and retrieved through its emission (estimated mean values after a finite number of repetitions), which is processed via a trained output layer. With this method, we show the phase precision limit following the standard quantum limit, the Heisenberg limit, and beyond.}
\label{FIG_setup}
\end{figure}

We define a quantum network as a collection of two-level quantum systems with random energies and all-to-all couplings as illustrated in Fig.~\ref{FIG_setup}.
In this paper, we utilise a bipartite (two-mode) state $|\psi_N\rangle$, which, after obtaining a phase information $\phi$ through the sample, interacts with the network.
We take all couplings to be energy-preserving, i.e., Josephson or Jaynes-Cummings type, such that the Hamiltonian is written as
\begin{eqnarray}\label{EQ_hamiltonian}
\hat H&=&\sum_j^{\mathcal{Q}} E_j \hat b_j^{\dagger} \hat b_j + \sum_{jj^{\prime}}^{\mathcal{Q}} C_{jj^{\prime}} \left( \hat b_j^{\dagger} \hat b_{j^{\prime}} + \hat b_{j^{\prime}}^{\dagger}\hat b_j \right) \nonumber \\
&&+\sum_{j}^{\mathcal{Q}} \sum_{k=1,2} W_{jk} \left( \hat a_k^{\dagger} \hat b_j + \hat b_j^{\dagger} \hat a_k \right),
\end{eqnarray}
where $E_j$ is the energy of the $j$th node, whose annihilation (creation) operator is denoted by $\hat b_j$ ($\hat b_j^{\dagger}$), $C_{jj^{\prime}}$ the coupling strengths between the QN nodes, and $W_{jk}$ stands for the coupling strength between the network node $j$ and an input mode $k$. 
We have used $\mathcal{Q}$ to indicate the number of nodes used in the QN.
We also consider other types of coupling, i.e., ultra strong non-energy preserving and cascading type -- see Appendix~\ref{App_othercouping} for detailed expressions.

As the input for the QN, we first consider NOON states, which we write as 
\begin{equation}\label{EQ_noon}
|\psi_N\rangle \equiv \frac{1}{\sqrt{2}}\left(|N0\rangle - |0N\rangle\right),
\end{equation}
where $N$ denotes the number of excitations.
The preparation of NOON states has been demonstrated experimentally, e.g., in Refs.~\cite{walther2004broglie,mitchell2004super,afek2010high}.
See also Ref.~\cite{creating} for their preparation, specifically in the form of Eq.~(\ref{EQ_noon}).
Note that a relative phase in the input state, such as the minus sign in Eq.~(\ref{EQ_noon}) is irrelevant for the present task. 
One can utilise other forms of resource state, e.g., we will also use maximally entangled states and classically correlated states later.
After passing one mode of the state through the sample, the state carrying the phase information reads $\left(|N0\rangle - \exp{(iN\phi)}|0N\rangle\right)/\sqrt{2}$.

For the dynamics of our setup, in addition to the coherent evolution corresponding to the Hamiltonian of Eq.~(\ref{EQ_hamiltonian}), we also consider the quantum network being subjected to noises.
In particular, we write the evolution of the density matrix $\rho$ of the whole system as follows:
\begin{equation}\label{EQ_evolution}
\rho(t+\Delta t)=\hat D(\Delta t)\: \hat U(\Delta t) \:[\rho(t)],
\end{equation}
where $\hat D$ and $\hat U$ denote the application of possible noise channels and unitary $\exp{(-i\hat H\Delta t/\hbar})$, respectively.
We take into account the energy decay $\hat D_{\text{dcy}}$, dephasing $\hat D_{\text{dph}}$, and depolarising $\hat D_{\text{dpl}}$ processes affecting the QN nodes.
See Appendix~\ref{APP_noises} for explicit expressions, where the strength of the noises is characterised by $\gamma_{\text{dcy}},\gamma_{\text{dph}},\gamma_{\text{dpl}}$, in units of energy.

We assume that it is possible to estimate the mean excitation numbers of the QN nodes and that they can be linearly combined with a set of tuneable weights. This corresponds to the action of an output layer of our network, producing an output signal:
\begin{equation}\label{EQ_fout}
f_{\text{out}}=\alpha_0+\alpha_1 \langle n_1\rangle+\alpha_2 \langle n_2\rangle + \cdots + \alpha_{\mathcal{Q}} \langle n_{\mathcal{Q}}\rangle,
\end{equation} 
where $\langle n_j \rangle=\text{tr}(\hat b_j^{\dagger}\hat b_j \rho(\tau))$ denotes the ideal mean excitation of the $j$th QN node at time $\tau$.
The coefficients, written in a vector form, ${\bm{\alpha}}\equiv(\alpha_0,\alpha_1,\alpha_2,\cdots,\alpha_{\mathcal{Q}})^T$ are trained such that the error of the output signal is minimised.
The training is performed with ridge regression (see Appendix~\ref{APP_training} for details).
Later, we show that our method also allows for time-integrated measurements (instead of time-resolved at time $\tau$), i.e., $f_{\text{out}}=\alpha_0+\alpha_1 \int \langle n_1\rangle dt / T+\alpha_2 \int \langle n_2\rangle dt / T+\cdots$, where $T$ is the measurement duration.

In experimental situations, mean excitation numbers are determined from a finite number of measurements and their optimal use is the subject of metrology.
To account for the deviation from the ideal values, which require an infinite number of measurements, we introduce a random error as follows
\begin{equation}\label{EQ_deviation}
\langle n_j \rangle = \langle n_j \rangle_{\text{ideal}} +\epsilon_j,
\end{equation}
where $\langle n_j \rangle_{\text{ideal}}=\text{tr}(\hat b_j^{\dagger}\hat b_j \rho(\tau))$ and $\epsilon_j$ is a random number normally distributed with zero mean and standard deviation of the mean (SDM) $\xi/2$.
By the central limit theorem it follows that $\xi \propto 1/\sqrt{M}$ (the SQL), where $M$ is the number of measurements.
We note that for $\xi \rightarrow0$, the simulations show that our method reproduces $\Delta \phi \rightarrow 0$ for phase estimation as it should be since this corresponds to $M \to \infty$.
Importantly, the addition of systematic errors in Eq.~(\ref{EQ_deviation}) has no effect on $\Delta \phi$, because the training procedure learns to overcome them.

\section{Results}

In what follows, we shall set the network parameters $(E_j,C_{jj^{\prime}},W_{jk})=(e_j,c_{jj^{\prime}},w_{jk})\hbar \Omega$, where the lowercase parameters are dimensionless and $\hbar \Omega$ has units of energy.
To simulate the imperfections in fabrication of the QN, we generate random parameters $(e_j,c_{jj^{\prime}},w_{jk})\in[0,1]$ that are uniformly distributed.
As the initial condition for the QN, we assume the experimentally sensible ground state for all the nodes, i.e., $|0\rangle^{\otimes \mathcal{Q}}$.

\subsection{Output signal}
Following the discussion from Refs.~\cite{giovannetti2004quantum,nagata2007beating}, we take the following function as the target output signal:
\begin{equation}\label{EQ_idealsignal}
I_{\text{ideal}}=\frac{1}{2}\left(1-\cos(N\phi)\right),
\end{equation}
where $N$ corresponds to the degree of NOON states used in the input. 
The function $I_{\text{ideal}}$ can show both super resolution due to its $N\phi$ dependence and super sensitivity, which we will demonstrate later.
For one realisation of the set of random network parameters $(e_j,c_{jj^{\prime}},w_{jk})$, the assessment of our method is conducted as follows.
For the training procedure, we generate $N_{\text{train}}$ random phases $\phi$, which lead to $N_{\text{train}}$ sets of $( \langle n_1\rangle, \langle n_2\rangle,\cdots,\langle n_{\mathcal{Q}}\rangle)$ from the realised QN, and in each set the mean values are estimated after $M$ measurements according to Eq.~(\ref{EQ_deviation}).
The coefficients $\bm{\alpha}$ (output layer) are trained using the $N_{\text{train}}$ training sets with ridge regression (see Appendix~\ref{APP_training}) such that the estimated output signal, 
\begin{equation}\label{EQ_Iestimate}
I_{\text{est}}=\alpha_0+\alpha_1 \langle n_1\rangle+\alpha_2 \langle n_2\rangle+\cdots+\alpha_{\mathcal{Q}} \langle n_{\mathcal{Q}}\rangle,
\end{equation}
is close to the ideal form in Eq.~(\ref{EQ_idealsignal}).
To test the trained output layer, we generate $N_{\text{test}}$ random phases, labelled $\phi_l$, which consequently give an estimated output signal $I_{\text{est},l}$.
We note that the training is performed only once, after which we obtain the coefficients ${\bm{\alpha}}$ that one can use to retrieve the estimated output of any phase $\phi_l$.

\begin{figure}[h]
\centering
\includegraphics[width=0.48\textwidth]{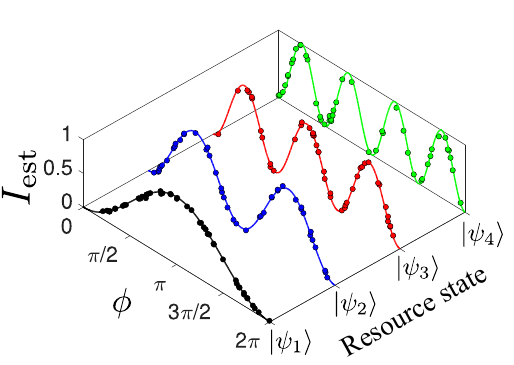}
\caption{Exemplary output signals showing super resolution. Different NOON states $|\psi_N\rangle$ were used as input for a quantum network of size 4. 
The corresponding solid curves indicate the ideal signals $I_{\text{ideal}}$. In all cases, the training and testing size is 10 and 50, respectively, with $\xi=0.01$.}
\label{FIG_Iout}
\end{figure}

We present in Fig.~\ref{FIG_Iout}, an exemplary demonstration of our method using one realisation of the network parameters, showing estimated output signals (filled circles) and the corresponding ideal ones (solid curves). 
We utilised NOON states $N=1$, $2$, $3$, and $4$. 
In all cases, we used $N_{\text{train}}=10$ and $N_{\text{test}}=50$, the measurement error $\xi=0.01$, and a quantum network of size 4 with evolution time $\tau=8/\Omega$.
Fig.~\ref{FIG_Iout} shows super resolution for higher $N$-NOON states, as seen from the $2\pi/N$ period of oscillations in the range $\phi \in [0,2\pi)$. 
It is also intuitive that the retrieval of the phase $\phi_{\text{est}}$ from $I_{\text{est}}$ has different accuracy for different $\phi$. 
In particular, the accuracy is best in the region of highest slope, e.g., $\phi=\pi/2N$ for the case of $N$-NOON state.
This will be quantified and demonstrated in more detail in the next section.

\subsection{Phase estimation}

We evaluate the phase from the output signal (for $N$-NOON state) as
\begin{equation}\label{EQ_phest}
\phi_{\text{est}}=\frac{1}{N} \arccos(1-2I_{\text{est}}).
\end{equation}
The measurement error $\epsilon_j$ in Eq.~(\ref{EQ_deviation}) can result in an output signal $I_{\text{est}}\notin [0,1]$ at extreme regions, e.g., $\phi\approx 0$ and $\pi/N$ in Fig.~\ref{FIG_Iout}, which further gives a complex $\phi_{\text{est}}$. 
To avoid these instances, we assign $\phi_{\text{est}}=0$ and $\phi_{\text{est}}=\pi/N$, for $I_{\text{est}} < 0$ and $I_{\text{est}} > 1$, respectively.
The error for the phase estimation task is quantified as follows: 
\begin{equation}\label{EQ_def_error}
\Delta \bar \phi_N=\sqrt{\sum_l^{N_{\text{test}}} \frac{(\phi_{\text{est},l}-\phi_{l})^2}{N_{\text{test}}(N_{\text{test}}-1)}}.
\end{equation}
Note that the bar notation indicates the testing is performed, where the phases $\phi_l$ are randomly generated over a range of values, e.g., $[0,\pi/N]$. 
For a particular value of the testing phase, i.e., $\phi_l=$ constant, the above expression reduces to the SDM of that phase, which we shall write simply as $\Delta \phi_N$ hereafter.

\begin{figure}[h]
\centering
\includegraphics[width=0.45\textwidth]{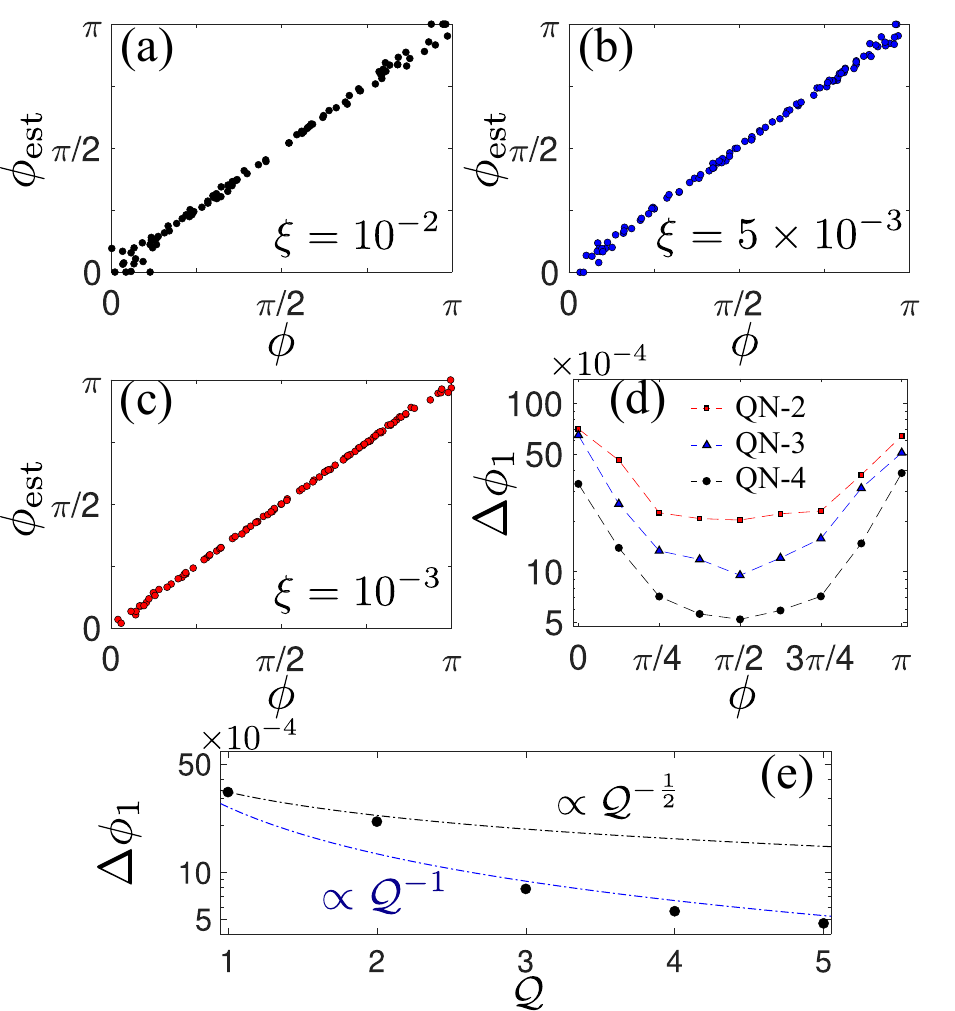}
\caption{Performance of phase estimation task. 
(a)-(c) Estimated phase vs ideal phase for different measurement error $\xi$, indicated on the bottom right of each panel. 
A QN with 4 nodes was used in these simulations. (d) SDM at different $\phi$ for different network sizes: (2, 3, 4) indicated by (squares, triangles, circles), respectively, and $\xi = 10^{-3}$. 
(e) $\Delta \phi_1$ at $\pi/2$ indicating the scaling with QN size.
Each SDM is averaged over 50 sets of realisations of the random network parameters. 
In all considered cases, for each realisation of the network parameters, $|\psi_1\rangle$ was used with 10 training and 100 testing sets.}
\label{FIG_phase}
\end{figure}

We demonstrate the estimated phase vs ideal phase in Fig.~\ref{FIG_phase} for different strengths of the measurement error: $\xi=10^{-2}$ (a), $5\times 10^{-3}$ (b), and $10^{-3}$ (c).
We have used a quantum network with 4 nodes that is evolved for $\tau=8/\Omega$, NOON state $|\psi_1\rangle$ as the input resource state, 10 training sets, and 100 testing sets.
It can be seen that $\phi_{\text{est}}$ deviates more from the ideal phase near $\phi=0$ and $\pi$ (low output slope, see Fig.~\ref{FIG_Iout}), while showing best accuracy at $\phi=\pi/2$. 
It is also clear that smaller measurement error $\xi$ produces less error in the estimated phase.

Fig.~\ref{FIG_phase}(d) compares the performance of different QN sizes, i.e., $\mathcal{Q}=2$ (squares), $3$ (triangles), and $4$ (circles) with $\tau=8/\Omega$.
NOON state $|\psi_1\rangle$ and a measurement error $\xi=10^{-3}$ were used in all three cases.
We realised 50 sets of the random network parameters, in each of which we used $N_{\text{train}}=10$ and $N_{\text{test}}=100$.
We computed the SDM $\Delta \phi_1$ (averaged over 50 sets of QN realisation) at different $\phi$.
This way, our method is not highly dependent on a particular realisation of the network parameters.
It is clear that more network nodes produce less error. 
Moreover, Fig.~\ref{FIG_phase}(d) not only indicates that the least error is found around $\phi=\pi/2$, but also how the error behaves as a function of $\phi$.
Additionally, Fig.~\ref{FIG_phase}(e) shows how the phase estimation error $\Delta \phi_1$ changes with respect to the QN size $\mathcal{Q}$. 
The scaling exceeds $\propto \mathcal{Q}^{-1/2}$, which is the SQL with respect to the number of QN nodes.

\begin{figure*}[t]
\centering
\includegraphics[width=0.9\textwidth]{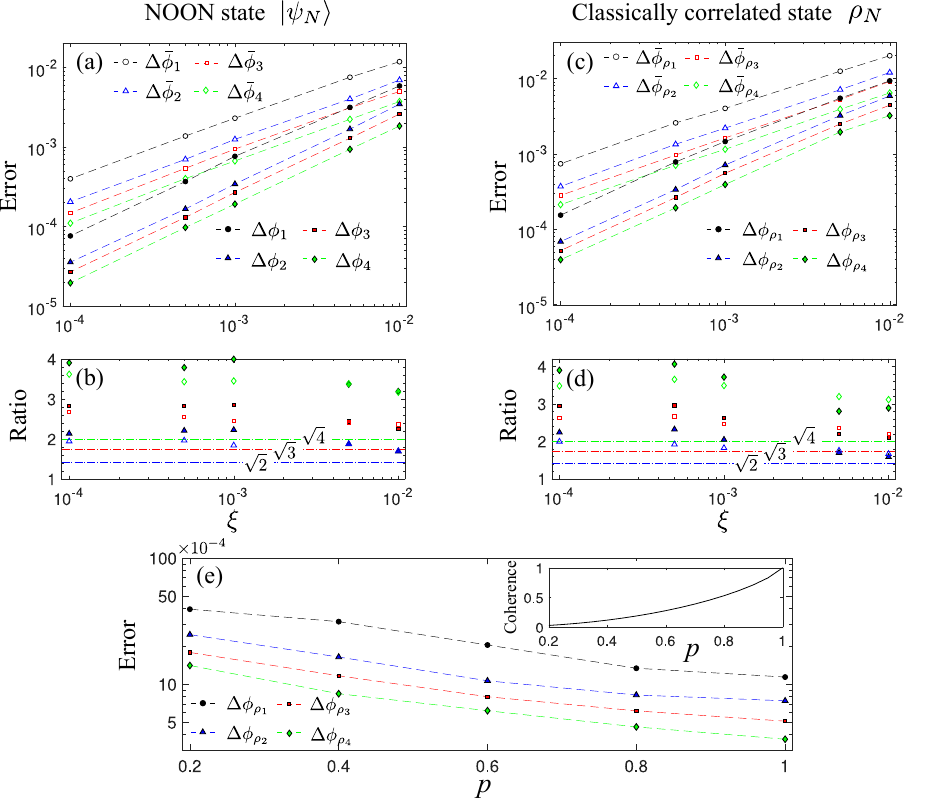}
\caption{Phase precision showing the standard quantum limit and Heisenberg limit. 
(a) Phase estimation errors $\Delta \bar \phi_N$ and the SDM at highest output slope $\Delta \phi_N$, plotted against the measurement error $\xi$ ($\propto 1/\sqrt{M}$). 
We have used $N$ to denote the use of $N$-NOON state. 
(b) The ratio of phase estimation errors, see Eq.~(\ref{EQ_ratio}). Empty triangles are for $\bar \eta_{12}$, empty squares for $\bar \eta_{13}$, and empty diamonds for $\bar \eta_{14}$. 
Also the corresponding filled ones: $\eta_{12}$ (triangles), $\eta_{13}$ (squares), and $\eta_{14}$ (diamonds). 
We report that $\Delta \bar \phi_N,\Delta \phi_N \propto 1/\sqrt{M}$ -- demonstrating the SQL limit, and $\Delta \bar \phi_N,\Delta \phi_N \propto 1/N$ -- demonstrating the HL limit. 
Panels (c) and (d) are the phase estimation errors and their ratios, respectively, using the classically correlated state $\rho_N$ in Eq.~(\ref{EQ_classcorstate}).
(e) Estimation errors using $\rho_N$ dephased in the Fock basis, modelled by multiplying the off-diagonal elements of $\rho_N$ with a positive coefficient $p\le1$.
The inset shows the amount of coherence in the state $\rho_N$ with respect to $p$.
In all cases, each data point represents an error evaluation of a trained output layer with 10 training and 100 testing sets, which is averaged over 50 different realisations of network parameters.}
\label{FIG_qadv}
\end{figure*}

For any to-be-measured phase $\phi$, one can train the QN with a ``shift", i.e., $I_{\text{ideal}}=\left(1-\cos(N(\phi+\theta))\right)/2$.
The shift $\theta$ is chosen such that the highest output slope is located around $\phi$, resulting in the minimum SDM of the phase. 
See a demonstration in Appendix~\ref{APP_highestoutputslope}. 
Hereafter, the SDM $\Delta \phi_N$ for any random $\phi$ will be assumed at the highest output slope.
Note that this procedure is all done in the data processing step (post measurements).
The next section quantifies and expands on the observations made about Fig.~\ref{FIG_phase}.

\subsection{Phase precision scaling} 

Here we demonstrate phase precision scaling, which follows the SQL, and HL by utilising higher degree $N$-NOON states.
For this purpose, we used a QN composed of 4 nodes, evolved for $\tau=12/\Omega$.
Fig.~\ref{FIG_qadv}(a) presents the phase estimation errors, both $\Delta \bar \phi_N$ and the SDM $\Delta \phi_N$ at the highest output slope, where $N$ indicates the use of $N$-NOON state.
These errors were plotted against the measurement error $\xi$, defined in Eq.~(\ref{EQ_deviation}).
We generated 50 different sets of realisations of the network parameters.
In each of these sets, we performed the training of the output layer with $N_{\text{train}}=10$ and the phase estimation error was tested with $N_{\text{test}}=100$. 
Each data point in Fig.~\ref{FIG_qadv}(a) is the average phase estimation error of the 50 different realisations. 
One can see that $\Delta \bar \phi_N,\Delta \phi_N \propto \xi$. 
As $\xi \propto 1/\sqrt{M}$, where $M$ denotes the number of measurements, it follows that $\Delta \bar \phi_N,\Delta \phi_N \propto 1/\sqrt{M}$, which is the SQL statement.

We now evaluate closer the use of higher $N$-NOON states for a better scaling option.
The argument goes as follows: for example, instead of having double the number of measurements with $|\psi_1\rangle$ one can utilise $|\psi_2\rangle$, and harness the quantum advantage for precision beyond the SQL.
Note that the comparison is made, where in both cases, one has the same number of photons passing through the sample.
For a more general scenario, in order to beat the SQL, one has to show that the phase error for $N$-NOON state ($N>1$) gives a ratio 
\begin{eqnarray}\label{EQ_ratio}
\bar \eta_{1N}&\equiv& \Delta \bar \phi_1/\Delta \bar \phi_N>\sqrt{N},
\end{eqnarray}
or for the case of SDM, $\eta_{1N}\equiv \Delta \phi_1/\Delta \phi_N>\sqrt{N}$.
Fig.~\ref{FIG_qadv}(b) shows the ratio $\bar \eta_{1N}$: empty triangles ($N=2$), squares ($3$), and diamonds ($4$), and the corresponding filled ones for $\eta_{1N}$.
Indeed, not only the ratios exceed the SQL scaling indicated by the dash-dotted lines, they also approach $\bar \eta_{1N}, \eta_{1N}\to N$ -- the HL scaling.

The quantum network platform allows for precision beyond the SQL even if the input resource states do not posses quantum correlations.
In this case, we consider states of the form
\begin{eqnarray}
\rho_N &= & \frac{1}{2}|\psi_N \rangle \langle \psi_N|+\frac{1}{2}|\tilde \psi_N \rangle \langle \tilde \psi_N| \label{EQ_classcorstate} \\
& = & \frac{1}{2} |+- \rangle \langle +- | + \frac{1}{2} |-+ \rangle \langle -+ |,  \label{EQ_CL_CORR}
\end{eqnarray}
where we introduced two-mode state $|\tilde \psi_N\rangle \equiv (|00\rangle-|NN\rangle)/\sqrt{2}$
and single-mode states $| \pm \rangle = (|0\rangle \pm |N\rangle)/\sqrt{2}$.
In order to understand the resources present in $\rho_N$, let us recall that quantum entanglement is a special type of quantum correlation present between quantum systems~\cite{RevModPhys.81.865}, and can be quantified, e.g., with negativity~\cite{negativity}.
A broader class of quantum correlations is known as quantum discord~\cite{discord1,discord2}.
It draws the border between quantum and classical correlations, and has been shown as a necessary ingredient for entanglement gain between mediated systems~\cite{streltsov2012quantum, chuan2012quantum,krisnanda2017revealing,krisnanda2020distribution}.
One can infer that not only the state $\rho_N$ is separable (not entangled) $E_{1:2}=0$, it also has zero quantum discord $D_{1|2}=D_{2|1}=0$, and hence, contains only classical correlations. 
This is apparent (without calculations) since $\rho_N$ can be written in a form that only requires orthogonal states for the subsystems, i.e., the $|\pm\rangle$ in Eq.~(\ref{EQ_CL_CORR}).
However, it has coherence (off-diagonal elements) when represented in the Fock basis.
Also, note that the state $\rho_N$ can be thought of as embedded in a two qubit space, with the levels given by $|0\rangle$ and $|N\rangle$.
This way, $\rho_N$ is simply an equal mixture of two Bell-like states.


With the state $\rho_N$, we performed phase estimation tasks, similar to those in Fig.~\ref{FIG_qadv}(a).
The results are plotted in Fig.~\ref{FIG_qadv}(c), where the parameters and notation are the same as in Fig.~\ref{FIG_qadv}(a).
Although the phase estimation errors are slightly higher than those in Fig.~\ref{FIG_qadv}(a), the scaling for higher $N$ still beats the SQL and approaches the HL, see Fig.~\ref{FIG_qadv}(d).
This finding opens up a new path of performing super sensitive phase measurements using resource states that do not have quantum correlations and are relatively easier to prepare.
We note that previous work conjectured the role of quantum discord in mixed state quantum metrology~\cite{modi2011quantum}. 
Our work extends this direction and presents metrology without quantum discord.
In this case, the coherence of the state $\rho_N$ plays an important role. 
In order to demonstrate this more closely, suppose the off-diagonal elements of the state $\rho_N$ are multiplied by a positive number $p\le 1$, which simulates dephasing in the Fock basis.
Complete dephasing is given when $p=0$, in which case the state, now completely diagonal, cannot carry the phase information $\phi$.
Phase estimation errors for different values of $p$ are plotted in Fig.~\ref{FIG_qadv}(e).
The inset shows how the coherence, quantified as $Sn(\rho_N(p=0))-Sn(\rho_N(p))$~\cite{baumgratz2014quantifying}, where $Sn(\rho)\equiv -\text{tr}(\rho \ln{\rho})$ denotes the von Neumann entropy, changes with respect to the variable $p$.
It is clear that states with larger coherence result in less phase estimation errors.


\subsection{Time-resolved and time-integrated processing} 

\begin{figure}[h]
\centering
\includegraphics[width=0.48\textwidth]{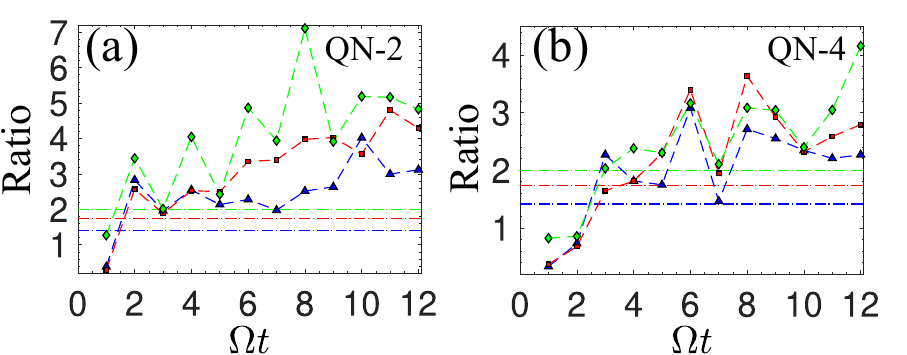}
\caption{SDM ratio $\eta_{1N}$ at different times for quantum network with $2$ (a) and $4$ (b) nodes. Notation as in Fig.~\ref{FIG_qadv}(b). At some times, the ratio $\eta_{1N}$ exceeds not only the SQL ($\sqrt{N}$) but also the HL (N). Although the ratio at some times for QN-$2$ is higher, the QN-$4$ produces less phase estimation errors because its performance for single-excitation NOON state (the reference for the ratio) is much better, see Fig. 3(d).}
\label{FIG_time}
\end{figure}

Thus far, we have considered processing mean values $\{\langle n_j\rangle\}$ at a particular time $\tau$. 
One might ask how the quantum advantage (beating the SQL) changes with respect to time. 
To answer this question, we present the SDM ratio $\eta_{1N}$ at different times in Fig.~\ref{FIG_time}, both using $\mathcal{Q}=2$ (a) and $4$ (b) network nodes. 
In both panels, the ratios are denoted as: $\eta_{12}$ (triangles), $\eta_{13}$ (squares), and $\eta_{14}$ (diamonds).
The corresponding SQL scaling thresholds are indicated by the dash-dotted lines: blue ($\sqrt{2}$), red ($\sqrt{3}$), and green ($\sqrt{4}$).
All the values $\eta_{1N}$ are obtained in the same way as in Fig.~\ref{FIG_qadv} for $\xi=10^{-3}$, i.e., with training size 10, testing size 100, and averaged over 50 realisations of the network parameters. 
It can be seen that it takes time for the quantum advantage to surface, especially for the case of QN-$4$, which involves more network nodes. 
This is intuitive since it requires time for the information to be embedded in the quantum network, even so in one with a bigger size. 
From Fig.~\ref{FIG_time}, the quantum advantage can reach higher values for QN-$2$ in the considered time span.
However, this does not mean that QN-$2$ performs better than QN-$4$, as the bigger size QN offers lower phase estimation errors $\Delta \phi_N$ (not shown).
We note that at some times, the quantum advantage exceeds even the HL scaling, e.g., at $\Omega t=8$ in Fig.~\ref{FIG_time}(a). 
This is inline with the prediction that interacting systems with multipartite couplings, as it is the case for the QN here, can indeed go beyond the HL~\cite{boixo2007generalized,choi2008bose,roy2008exponentially}.
We also performed similar analysis using different coupling types and other entangled states as input, see Section~\ref{APP_othercouplingresults}.

A more experimentally friendly option considers processing time-integrated mean values from the QN nodes to form the estimated output signal, i.e., $I_{\text{est}}=\alpha_0+\alpha_1 \int \langle n_1\rangle dt / T+\alpha_2 \int \langle n_2\rangle dt / T+\cdots$, from which the estimated phase $\phi_{\text{est}}$ is calculated.
To exemplify this point, we utilised the time-integrated mean values for the scenario in Fig.~\ref{FIG_time}, where the measurement was conducted from $\Omega t=10.75$ to $11.25$.
For the case of QN-$2$ the resulting ratios are $\eta_{12}\approx2.1$, $\eta_{13}\approx3.7$, and $\eta_{14}\approx4.1$, whereas for QN-$4$ they are given by $2.6$, $3.2$, and $3.5$, respectively.
In all cases, the quantum network approach offers quantum advantage, i.e., the ratio beyond the SQL, and often beyond HL.

\subsection{Noises} 

Here we shall investigate the role of noise, i.e., energy decay, dephasing, and depolarising channels affecting the QN nodes.
Let us consider QN-$4$, where the bigger network size is likely to cause more disturbance from noise, which is evolved for a time $\tau=6/\Omega$.
To scrutinise each source of noise, we shall study the application of the noise channels: energy decay $\hat D_{\text{dcy}}$, dephasing $\hat D_{\text{dph}}$, and depolarising $\hat D_{\text{dpl}}$ separately. 
The strength of these channels are characterised, respectively, by $\gamma_{\text{dcy}}$, $\gamma_{\text{dph}}$, and $\gamma_{\text{dpl}}$ in units of energy (see Appendix~\ref{APP_noises} for details).
For simplicity, we have assumed the same noise strength for all the QN nodes.

\begin{figure}[h]
\centering
\includegraphics[width=0.45\textwidth]{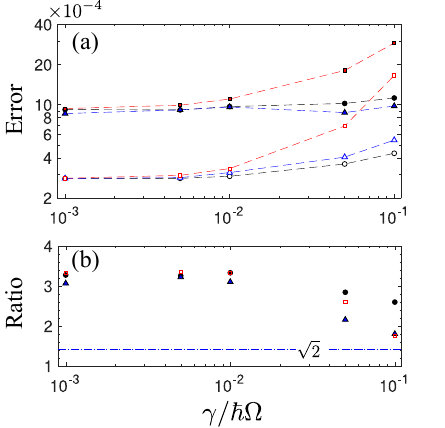}
\caption{(a) SDM of highest slope $\Delta \phi_1$ (filled symbols) and $\Delta \phi_2$ (empty symbols) in the presence of energy decay (circles), dephasing (triangles), and depolarising (squares) noise. The axis $\gamma/\hbar \Omega$ refers to the ratio of the strength of respective noise to the energy unit of the quantum network parameters. The simulations were conducted averaging over 50 realisations of network parameters, with 10 training and 100 testing sets. A QN with size $4$ was used, and $\tau=6/\Omega$. (b) The ratio $\eta_{12}$ of the SDMs in panel (a) for the case of energy decay (circles), dephasing (triangles), and depolarising (squares) noise. Quantum advantage is present for all these noise strengths.}
\label{FIG_noise}
\end{figure}

In Fig.~\ref{FIG_noise}(a) we present the average SDM $\Delta \phi_N$ against the noise-to-QN energy ratio $\gamma/\hbar \Omega$.
The filled (empty) symbols indicate the use of $|\psi_1\rangle$ ($|\psi_2\rangle$), with the shapes corresponding to the application of energy decay (circles), dephasing (triangles), and depolarising channel (squares).
Similar to the scenario in Fig.~\ref{FIG_time}(b), the average is taken over 50 different realisations of the network parameters, each with 10 training and 100 testing sets.
It is apparent that the use of higher $N$-NOON state, $|\psi_2\rangle$ in Fig.~\ref{FIG_noise}(a), is affected more severely by all the noises.
This is expected as higher degree NOON states possess more excitations.
We also note that depolarising noise has the worst effect on the phase estimation error. 
This channel corresponds to mixing the state of the QN nodes with a maximally mixed state, rendering part of the network ``capacity" useless for information embedding.
Note that up to $\gamma/\hbar \Omega=10^{-2}$, the effects from all the noises are minute. 
The error ratio $\eta_{12}$ is plotted in Fig.~\ref{FIG_noise}(b) where the affecting noise is energy decay (circles), dephasing (triangles), and depolarising (squares).
It can be seen that even in the situation where $\gamma/\hbar \Omega=0.1$, the quantum advantage still persists.

\subsection{Beating SQL with other coupling mechanisms and resource states}\label{APP_othercouplingresults}

Now we demonstrate that our method is not limited to the type of coupling involved between the quantum systems.
We shall vary the coupling type between the input and QN nodes, which is responsible for embedding the phase information into the QN.
In particular, we consider: energy-preserving type coupling (EP), also known as Jaynes-Cummings or Josephson coupling; ultra strong coupling; and cascading~\cite{gardiner1993driving,carmichael1993quantum} of the input into the QN.
The model for ultra strong coupling is similar to the EP, where the last term in the Hamiltonian of Eq.~(\ref{EQ_hamiltonian}) changes to $\sum_{j}^{\mathcal{Q}} \sum_{k=1,2} W_{jk} \left( \hat a_k^{\dagger} \hat b_j + \hat b_j^{\dagger} \hat a_k +\hat a_k \hat b_j + \hat b_j^{\dagger} \hat a_k^{\dagger} \right)$.
The addition of the last two terms in the summation indicates strong coherent interactions allowing for simultaneous annihilation and creation of excitations in the input party $k$ and QN node $j$.
For the cascading formalism, see Appendix~\ref{App_othercouping} for details.
Similarly, it involves an input-network coupling coefficient $W_{jk}$ in energy unit.
Also, in this formalism the input $k$ experiences an energy decay characterised by a coefficient $\sum_j W_{jk}^2/\gamma$, where $\gamma$ denotes a constant decay strength for the QN nodes.

\begin{figure}[t]
\centering
\includegraphics[width=0.45\textwidth]{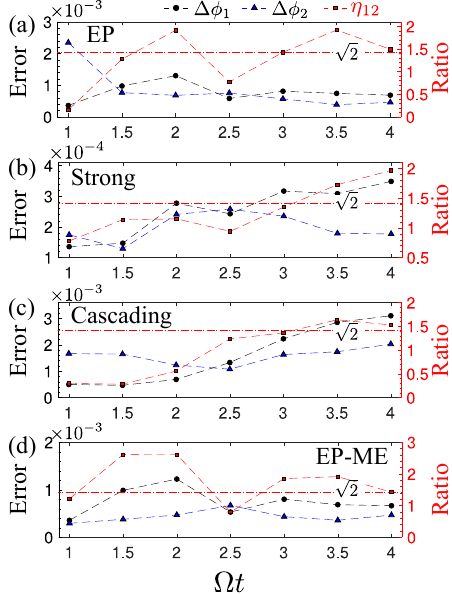}
\caption{Comparison of different input-network coupling mechanisms and resource states. (a) Energy-preserving coupling (Jaynes-Cummings or Josephson). (b) Ultra strong coherent interactions (note different vertical scale). (c) Cascading of input into QN nodes. (d) Energy-preserving coupling with maximally entangled states as the input. In all the panels, SDMs $\Delta \phi_1$ and $\Delta \phi_2$ are plotted for different evolution times. The right axis indicates the ratio $\eta_{12}=\Delta \phi_1/\Delta \phi_2$. All the considered cases are capable of beating the SQL threshold (red dash-dotted lines).}
\label{FIG_diffcoupling}
\end{figure}

\begin{figure*}[t]
\centering
\includegraphics[width=0.9\textwidth]{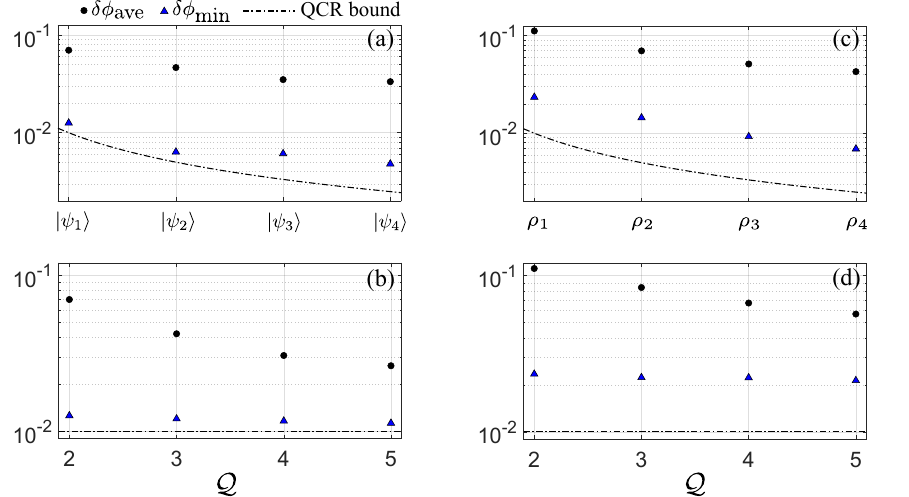}
\caption{Comparison of the phase measurement errors using a quantum neuromorphic platform to the quantum Cram\'{e}r-Rao bound.
With respect to $10^3$ realisations of the quantum network parameters, the average and minimum phase measurement errors are denoted by the black circles and blue triangles, respectively. 
The errors are plotted against: (a) different NOON states $|\psi_N\rangle$ using QN of size $\mathcal{Q}=2$; (b) QN sizes using $|\psi_1\rangle$; (c) different mixed states $\rho_N$ using QN of size 2; and (d) different QN sizes using $\rho_1$.
In all panels, the quantum Cram\'{e}r-Rao bound is indicated by the black dashed-dotted curve or line, where the number of repetitions is $M=10^4$.
The testing was performed at the highest output slope.
}
\label{FIG_qcr_bound}
\end{figure*}

Fig.~\ref{FIG_diffcoupling} shows the comparison between the EP (a), ultra strong (b), and cascading (c) coupling.
In all cases we used a QN with size 3, measurement error $\xi=10^{-3}$, and 50 realisations of the network parameters.
At each time $\Omega t$, we performed the training with 10 sets and the testing with 100 sets.
For Fig.~\ref{FIG_diffcoupling}(c) we set the decay $\gamma=\hbar \Omega$.
It is worth noting that the ultra strong type coupling produces lesser errors. 
This is partly because the ultra strong interactions allow for faster information embedding, i.e., the information spreads and occupies the capacity or Hilbert space of the QN more (with the addition of the terms $\hat a_k \hat b_j + \hat b_j^{\dagger} \hat a_k^{\dagger}$).
The cascading coupling has an advantage that the phase information travels in one way (into the QN), unlike the EP and the ultra strong couplings considered in Fig.~\ref{FIG_diffcoupling} where the evolution is coherent and part of the information travels back and forth between the input and the QN. 
Note that with a decay coefficient $\gamma=\hbar \Omega$, the cascading coupling is capable of producing nonclassical precision with estimated errors in the order $\sim 10^{-3}$.

Finally, the results in Fig.~\ref{FIG_diffcoupling}(d) were obtained with energy-preserving coupling where we used maximally entangled states as the resource (EP-ME). In this case, $|\psi_1\rangle=(|10\rangle+|01\rangle)/\sqrt{2}$ and $|\psi_2\rangle=(|20\rangle+|11\rangle+|02\rangle)/\sqrt{3}$. 
For the latter, after passing the sample, the input state reads $(|20\rangle+ \exp(i\phi)|11\rangle+ \exp(i2\phi)|02\rangle)/\sqrt{3}$. 
This state still carries $2\phi$ dependence, and the QN processing produces precision beating the SQL. 
We note that for EP-ME in Fig.~\ref{FIG_diffcoupling}(d), we used a slightly modified output model, i.e., $I_{\text{est}}=\alpha_0+\alpha_1\langle n_1\rangle+\alpha_2\langle n_2\rangle+\alpha_3\langle n_3\rangle+\alpha_4\langle n_1\rangle\langle n_2\rangle+\alpha_5\langle n_2\rangle\langle n_3\rangle+\alpha_6\langle n_3\rangle\langle n_1\rangle+\alpha_7\langle n_1\rangle\langle n_2\rangle\langle n_3\rangle$, where the coefficients were trained with ridge regression.
This way, it does not require extra measurements (i.e., only the mean values $\{\langle n_j\rangle\}$, as used in Eq.~(\ref{EQ_Iestimate})).
We note that in all panels of Fig.~\ref{FIG_diffcoupling}, the ratio $\eta_{12}$ exceeds the SQL threshold given by the red dash-dotted lines.

\subsection{The quantum Cram\'{e}r-Rao bound}

In quantum metrology, the minimum phase measurement error follows the so-called quantum Cram\'{e}r-Rao (QCR) bound:
\begin{equation}
    \delta \phi \ge \frac{1}{\sqrt{MF_q(\rho)}},
\end{equation}
where $M$ is the number of repetitions (measurements) and $F_q(\rho)$ is the quantum Fisher information (QFI) of a quantum state $\rho$.
Note that $\delta \phi$ is the standard deviation of the measurement error, different from that of Eq.~(\ref{EQ_def_error}) by $\sqrt{N_{\text{test}}}$.
The QFI of a general state $\rho$ can be evaluated following the expression given in Refs.~\cite{zhang2013quantum,liu2013phase}.
In the present case, we have $F_q(|\psi_N\rangle \langle \psi_N|)=F_q(\rho_N)=N^2$, which gives the QCR bound: $\delta \phi \ge 1/(\sqrt{M}N)$.

To compare our scheme to the QCR bound, we slightly modify the model for the measurement of the QN emission, previously described in Eq.~(\ref{EQ_deviation}), to explicitly take into account the number of repetitions $M$.
Given the ideal mean excitation of the $j$th QN node $\langle n_j \rangle_{\text{ideal}}$, $M$ random numbers are generated, labelled $\mu_m\in[0,1]$.
New values ($0$ or $1$) are assigned as follows: (i) $\tilde \mu_m=0$ if $\mu_m \ge \langle n_j \rangle_{\text{ideal}}$ and (ii) $\tilde \mu_m=1$ if $\mu_m < \langle n_j \rangle_{\text{ideal}}$.
This way, $\tilde \mu_m$s mimic real experimental data, the average of which converges to $\langle n_j \rangle_{\text{ideal}}$, with a standard deviation of the mean $\propto 1/\sqrt{M}$.

The phase measurement errors (standard deviation) are plotted in Fig.~\ref{FIG_qcr_bound}. 
The resource states in panels (a) and (b) are NOON states $|\psi_N\rangle$, whereas those in panels (c) and (d) are mixed states $\rho_N$.
The performance is plotted against the degree of resource state $N$ in panels (a) and (c), as well as against the number of QN nodes $\mathcal{Q}$ in panels (b) and (d).
The QCR bound with $M=10^4$ is plotted as a dashed-dotted curve or line in all the panels.
It can be seen that the minimum standard deviation $\delta \phi_{\text{min}}$ is close to the QCR bound, which indicates that there exists a set of QN parameters allowing for near-saturation performance. 
This way, one can think of the QN as a measuring and processing device to efficiently extract information from the phase-encoded input state.
In principle, one may perform a more rigorous and precise parameter search algorithm to find a better set of parameters for even smaller $\delta \phi_{\text{min}}$.
Here we simply take $10^3$ random realisations of the QN parameters, with evolution time $\tau=12/\Omega$.
It is expected that the performance of NOON states is better than the mixed states, despite having the same QFI.
Here, the scheme might benefit from a more complex QN architecture or a proper parameter search algorithm.
We also note that both $\delta \phi_{\text{ave}}$ and $\delta \phi_{\text{min}}$ are better for higher degree $N$ of the resource states and number of QN nodes $\mathcal{Q}$.

\section{Conclusion}

We have presented a platform for phase estimation tasks, based on a quantum network approach.
It consists of three main elements: (1) a resource state carrying a phase information as input; (2) a quantum network, which is made of a collection of randomly connected quantum systems (the nodes), acting as a quantum processing device; and (3) an output layer, which combines the emission or measurement results from the network nodes, and produces the final output.
The training is performed in the output layer with ridge regression such that the error of the target output is minimised.
The reported precision scales better than the standard quantum limit, and even the Heisenberg limit -- owing to the interacting nature of the quantum network.
We have shown that this is possible even with classically correlated states as input owing to quantum coherence.

Our proposed platform is versatile, i.e., it is applicable for different types of coupling between the input and quantum network: the explicit calculations covered Jaynes-Cummings or Josephson; ultra strong; and cascading coupling.
It also allows for both time-resolved and time-integrated processing of the network emissions.  
We show that the resulting quantum advantage is robust against energy decay, dephasing, and depolarising noises. 
One can further explore other forms of resource states as input or other types of coupling between the quantum systems involved.
This makes our platform attractive for a wide range of physical implementations.

\section*{Acknowledgment}
T.K., S.G., and T.C.H.L. acknowledge the support by the Singapore Ministry of Education under its AcRF Tier 2 grant MOE2019-T2-1-004. 
T.P. is supported by the Polish National Agency for Academic Exchange NAWA Project No. PPN/PPO/2018/1/00007/U/00001.

\newpage
\appendix

\section{Other types of input-network coupling}\label{App_othercouping}

For ultra strong coupling between the input parties and the quantum network nodes, the evolution is written as in Eq.~(\ref{EQ_evolution}) in the main text where the input-network coupling term in the Hamiltonian of Eq.~(\ref{EQ_hamiltonian}) now reads
\begin{equation}\label{EQ_hamiltonianstrong}
\sum_{j}^{\mathcal{Q}} \sum_{k=1,2} W_{jk} \left( \hat a_k^{\dagger} \hat b_j + \hat b_j^{\dagger} \hat a_k +\hat a_k \hat b_j + \hat b_j^{\dagger} \hat a_k^{\dagger} \right).\nonumber 
\end{equation}
The ultra strong coupling allows for simultaneous annihilation and creation of excitations, i.e., the terms $\hat a_k \hat b_j + \hat b_j^{\dagger} \hat a_k^{\dagger}$.

We also consider an input-network coupling following the cascading formalism~\cite{gardiner1993driving,carmichael1993quantum}.
The Hamiltonian in this case reads
\begin{equation}
\hat H=\sum_j^{\mathcal{Q}} E_j \hat b_j^{\dagger} \hat b_j + \sum_{jj^{\prime}}^{\mathcal{Q}} C_{jj^{\prime}} \left( \hat b_j^{\dagger} \hat b_{j^{\prime}} + \hat b_{j^{\prime}}^{\dagger}\hat b_j \right),
\end{equation}
and the cascading master equation has the following structure:
\begin{eqnarray}
\rho(t+\Delta t)&=&\rho(t)+\frac{\Delta t}{\hbar} \Big( -i [\hat H,\rho(t)] \nonumber \\
&&+\sum_j^{\mathcal{Q}} \frac{\gamma}{2} \mathcal{L}(\rho(t),\hat b_j) \nonumber \\
&&+\sum_{j}^{\mathcal{Q}} \sum_{k=1,2} W_{jk} [\hat a_k\rho(t),\hat b_j^{\dagger}]+[\hat b_j,\rho(t)\hat a_k^{\dagger}] \nonumber \\
&&+\sum_{k=1,2} \frac{\chi_k}{2} \mathcal{L}(\rho(t),\hat a_k) \Big),
\end{eqnarray}
where we have used $\mathcal{L}(\rho,\hat O)\equiv 2\hat O \rho \hat O^{\dagger}-\hat O^{\dagger}\hat O \rho-\rho \hat O^{\dagger}\hat O$ and $\chi_k=\sum_j^{\mathcal{Q}} W_{jk}^2/\gamma$.
In this formalism, $\gamma$ denotes the energy decay of the network nodes, while $\chi_k$ is that of the input party $k$.
The input-network coupling strength is characterised by $W_{jk}$.

\section{Noise channels}\label{APP_noises}

Here we provide detailed expressions for the energy decay, dephasing, and depolarising channels.
These channels are applied on the quantum network nodes, which are two-level quantum systems.
See Ref.~\cite{nielsenchuang} for a review.

The energy decay channel models the decay of excitations, i.e., from excited state to ground state over time.
It is denoted by the operator $\hat D_{\text{dcy}} (\Delta t)$, which is defined as follows:
\begin{eqnarray}
\rho(t+\Delta t)&=&\hat D_{\text{dcy}} (\Delta t) [\rho(t)] \\
&=&\sum_j^{\mathcal{Q}} \hat K_{j,1}\rho(t) \hat K_{j,1}^{\dagger}+\hat K_{j,2}\rho(t) \hat K_{j,2}^{\dagger},\nonumber 
\end{eqnarray}
where 
\begin{eqnarray}
\hat K_{j,1}&=&\left[\begin{array}{cc} 1 & 0 \\ 0 & \sqrt{1-\frac{\gamma_{\text{dcy}}}{\hbar} \Delta t} \end{array}\right],\nonumber \\
\hat K_{j,2}&=&\left[\begin{array}{cc} 0 & \sqrt{\frac{\gamma_{\text{dcy}}}{\hbar} \Delta t} \\ 0 & 0 \end{array}\right],
\end{eqnarray}
are operators applied on the $j$th node.
For simplicity, we have assumed that the decay strength $\gamma_{\text{dcy}}$ is the same for all the nodes.

Quantum states may loose their coherence, i.e., the decay of superposition (off-diagonal elements of $\rho(t)$).
In this case, the dephasing operator $\hat D_{\text{dph}} (\Delta t)$ is defined as:
\begin{eqnarray}
\rho(t+\Delta t)&=&\hat D_{\text{dph}} (\Delta t) [\rho(t)] \\
&=&\sum_j^{\mathcal{Q}} (1-\frac{\gamma_{\text{dph}}}{2\hbar} \Delta t)\rho(t)+\frac{\gamma_{\text{dph}} }{2\hbar}\Delta t \: \hat \sigma^z_{j}\rho(t) {\hat \sigma^z_{j}}, \nonumber 
\end{eqnarray}
where the dephasing is in the $\hat \sigma^z$ basis. 
The strength of this channel is characterised by $\gamma_{\text{dph}}$, which is assumed uniform.

Last, experimental conditions can also result in depolarisation of the quantum state $\rho(t)$, i.e., mixing with white noise.
This is modelled by $\hat D_{\text{dpl}} (\Delta t)$ defined as:
\begin{eqnarray}
\rho(t+\Delta t)&=&\hat D_{\text{dpl}} (\Delta t) [\rho(t)] \\
&=&\sum_j^{\mathcal{Q}} (1-\frac{\gamma_{\text{dpl}}}{\hbar} \Delta t)\rho(t) \nonumber \\
&&+\sum_j^{\mathcal{Q}} \sum_{\nu=x,y,z}\frac{\gamma_{\text{dpl}}}{3\hbar} \Delta t \: \hat \sigma^{\nu}_{j}\rho(t) {\hat \sigma^{\nu}_{j}}, \nonumber 
\end{eqnarray}
where $\gamma_{\text{dpl}}$ indicates the strength.

We have used Pauli matrices with the subscript $j$ denoting the application on the $j$th QN node:
\begin{eqnarray}
\hat \sigma^x_{j}&=&\left[\begin{array}{cc} 0 & 1 \\ 1 & 0 \end{array}\right],\:\hat \sigma^y_{j}=\left[\begin{array}{cc} 0 & -i \\ i & 0 \end{array}\right],\:\hat \sigma^z_{j}=\left[\begin{array}{cc} 1 & 0 \\ 0 & -1 \end{array}\right].
\end{eqnarray}

\section{Training of output layer}\label{APP_training}

Here we describe the training of the output layer, i.e., the coefficients $\bm \alpha \equiv (\alpha_0,\alpha_1,\alpha_2,\cdots,\alpha_{\mathcal{Q}})^T$ in Eq.~(\ref{EQ_Iestimate}).
Consider $N_{\text{train}}$ number of training sets, each of which consists of a known random phase, labelled $\phi_i$, and the resulting mean values from the QN $(\langle n_{1,i} \rangle,\langle n_{2,i} \rangle,\cdots,\langle n_{\mathcal{Q},i} \rangle)$ with a measurement error $\xi$, where the subscript denotes the $i$th training set. 
From the training sets, the coefficients $\bm \alpha$ are obtained as follows:
\begin{equation}
\bm \alpha= (\bm X^T \bm X + \lambda \openone)^{-1} \bm X^T \bm Y,
\end{equation}
where
\begin{eqnarray}
\bm X&=&\left[ \begin{array}{cccccc} 1 & \langle n_{1,1} \rangle & \langle n_{2,1} \rangle &\cdots & \langle n_{\mathcal{Q},1} \rangle \\ 
1 & \langle n_{1,2} \rangle & \langle n_{2,2} \rangle &\cdots & \langle n_{\mathcal{Q},2} \rangle \\
\vdots & \vdots  & \vdots &\ddots& \vdots  \\
1 & \langle n_{1,N_{\text{train}}} \rangle & \langle n_{2,N_{\text{train}}} \rangle &\cdots & \langle n_{\mathcal{Q},N_{\text{train}}} \rangle \end{array} \right]
\end{eqnarray}
and 
\begin{eqnarray}
\bm Y&=&\frac{1}{2}\left[ \begin{array}{c}  1-\cos(N\phi_1)\\ 1-\cos(N\phi_2) \\ \vdots \\ 1-\cos(N\phi_{N_{\text{train}}}) \end{array}\right].
\end{eqnarray}
The constant value $\lambda>0$ is known as the ridge parameter.

Once we obtain the coefficients $\bm \alpha$, we can calculate the output signal, given a set of mean values from experiments, via $I_{\text{est}}=\alpha_0+\alpha_1 \langle n_1 \rangle +\alpha_2 \langle n_2 \rangle +\cdots+\alpha_{\mathcal{Q}} \langle n_{\mathcal{Q}} \rangle$.
Next, from the output signal one can retrieve the phase $\phi_{\text{est}}$ via 
\begin{eqnarray}\label{EQ_phasere}
\phi_{\text{est}}&=&\Bigg\{ \begin{array}{l} 0,  \:\:I_{\text{est}} <0 \\ 
\frac{1}{N} \arccos(1-2I_{\text{est}}),  \:\:I_{\text{est}} \in [0,1] \\
\pi/N,  \:\:I_{\text{est}} >1. \end{array} 
\end{eqnarray}

For testing, we consider $N_{\text{test}}$ sets, each with a testing phase $\phi_l$ and the corresponding mean values $(\langle n_{1,l} \rangle,\langle n_{2,l} \rangle,\cdots,\langle n_{\mathcal{Q},l} \rangle)$, taking into account the measurement error $\xi$.
To quantify the phase estimation error, we use
\begin{equation}
\Delta \bar \phi_N=\sqrt{\sum_l^{N_{\text{test}}} \frac{(\phi_{\text{est},l}-\phi_{l})^2}{N_{\text{test}}(N_{\text{test}}-1)}},
\end{equation}
where $\phi_{\text{est},l}$ is calculated as in Eq.~(\ref{EQ_phasere}) and the subscript $N$ indicates the use of $N$-NOON state as a resource.

\begin{figure}[h]
\centering
\includegraphics[width=0.48\textwidth]{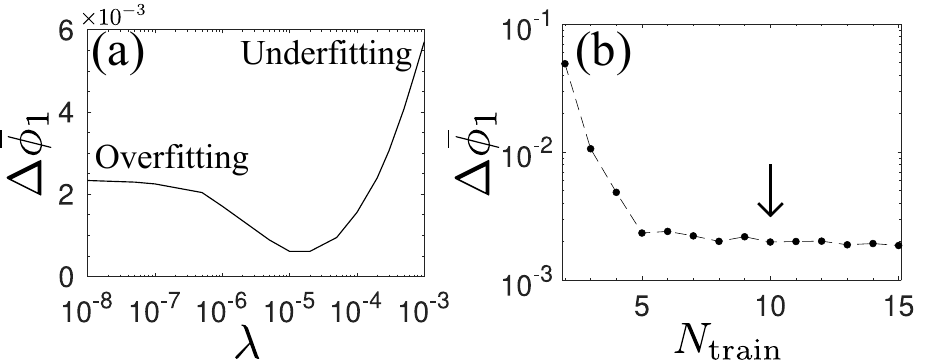}
\caption{Phase estimation error vs ridge parameter (a) and number of training sets (b). The arrow indicates the chosen $N_{\text{train}}$ in this paper.}
\label{FIG_training}
\end{figure}

We show the role of the ridge parameter $\lambda$ by considering the exemplary task presented in Fig.~\ref{FIG_phase}(c) where $\xi=10^{-3}$. 
In this case, we used 10 training sets and 100 testing sets. 
Fig.~\ref{FIG_training}(a) demonstrates the testing error against the ridge parameter $\lambda$. 
It is clear that the case where $\lambda \rightarrow 0$ (linear regression limit) leads to overfitting of the training data, i.e., including the measurement error, and thus overlooking the real behaviour.
On the other hand, higher value of $\lambda$ leads to underfitting. 
The tradeoff results in a minimum error, at $\lambda_{\text{min}}$.
The phase estimation errors reported in this paper follow this minimum error.

The number of training sets is another factor affecting the precision. 
To illustrate this, we plot in Fig.~\ref{FIG_training}(b) the phase estimation error against the number of training sets $N_{\text{train}}$.
We have used a QN with size 4, $\xi=10^{-3}$, and $N_{\text{test}}=100$.
Each point is averaged over 50 realisations of the network parameters.
It can be seen that $\Delta \bar \phi_1$ decreases with $N_{\text{train}}$. 
We have chosen $N_{\text{train}}=10$ throughout this paper.

\section{Standard deviation of the mean at highest output slope}\label{APP_highestoutputslope}

\begin{figure*}[t]
\centering
\includegraphics[width=0.9\textwidth]{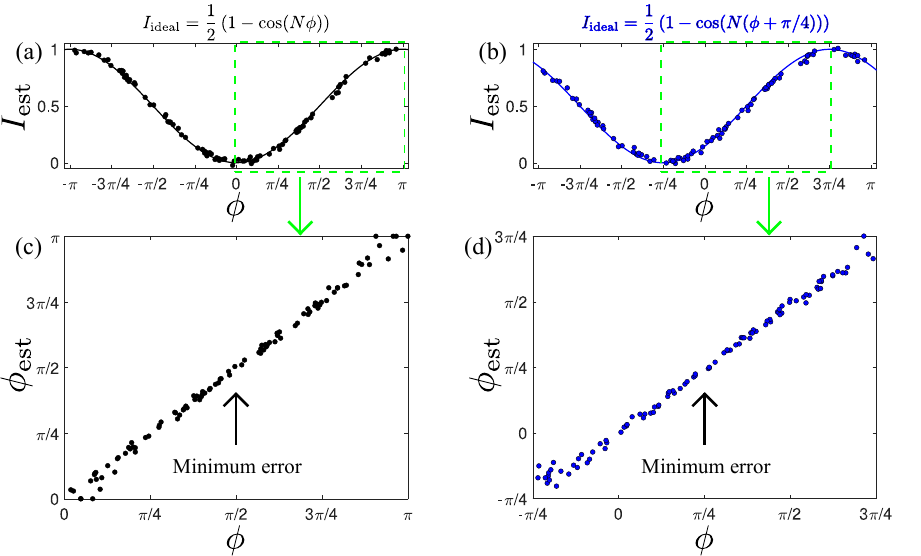}
\caption{Minimising the SDM with shifted output signal. The same training set and testing set are used for different target output $\theta=0$ (a) and $\pi/4$ (b). The corresponding phase estimation tasks in the range indicated by the green dashed boxes are plotted in (c) and (d), respectively. The minimum error shifts from $\theta=0$ (a) to $\pi/4$ (b).}
\label{FIG_mindphi}
\end{figure*}

For any constant phase $\phi$, one quantifies the phase estimation error with SDM $\Delta \phi_N$.
Here we demonstrate the advantage of our platform in the training step.
In particular, we consider a ``shifted" ideal output signal:
\begin{equation}\label{EQ_shiftedideal}
I_{\text{ideal}}=\frac{1}{2}\left(1-\cos(N(\phi+\theta))\right).
\end{equation}
The phase shift $\theta$ is chosen such that the SDM of the phase $\Delta \phi_N$ is minimum.
This corresponds to the phase $\phi$ situated at the highest slope of the output signal of Eq.~(\ref{EQ_shiftedideal}). 

To illustrate this, we plotted the estimated output signal for both $\theta=0$ and $\pi/4$ in Fig.~\ref{FIG_mindphi} panels (a) and (b), respectively.
We used the same training set ($N_{\text{train}}=10$) and testing set ($N_{\text{test}}=100$) in both panels.
Also, we used $\mathcal{Q}=4$ with $\tau=8/\Omega$ and $\xi=10^{-2}$.
The highest output slope shifts from $\phi=\pi/2$ in (a) to $\pi/4$ in (b).
Phase estimation in the range indicated by the green dashed boxes are plotted in panels (c) and (d).
It can be seen that the minimum error is located at the highest output slope.
In the data processing, one simply varies the phase shift $\theta$ until $\Delta \phi_N$ is minimised.

\end{document}